\documentclass[prl,aps,showpacs,onecolumn,floatfix]{revtex4}
\usepackage{epsfig} \usepackage{graphics} \usepackage{bm}
\usepackage{amssymb}
\usepackage{graphicx}
\begin{document}

\preprint{Lebed-2020}

\title{BREAKDOWN OF THE EINSTEIN'S EQUIVALENCE PRINCIPLE
FOR A QUANTUM BODY}

\author{ANDREI G. LEBED}

\affiliation{Department of Physics, University of Arizona, 1118 E. 4th Street,\\
Tucson, Arizona 85721, USA and\\
L.D. Landau Institute for Theoretical Physics, RAS, 2 Kosygina Street,\\
Moscow 117334, Russia\\
lebed@physics.arizona.edu}

\begin{abstract}
We review our recent theoretical results about inequivalence
between passive gravitational mass and energy for a composite
quantum body at a macroscopic level. In particular, we consider
macroscopic ensembles of the simplest composite quantum bodies -
hydrogen atoms. Our results are as follows. For the most
ensembles, the Einstein's Equivalence Principle is valid. On the
other hand, we discuss that for some special quantum ensembles -
ensembles of the coherent superpositions of the stationary quantum
states in the hydrogen atoms (which we call Gravitational demons)
- the Equivalence Principle between passive gravitational mass and
energy is broken. We show that, for such superpositions, the
expectation values of passive gravitational masses are not related
to the expectation values of energies by the famous Einstein's
equation, i.e, $m_g \neq \frac{E}{c^2}$. Possible experiments at
the Earth's laboratories are briefly discussed, in contrast to the
numerous attempts and projects to discover the possible breakdown
of the Einstein's Equivalence Principle during the space missions.
\end{abstract}

\keywords{Equivalence principle; Mass-energy equivalence; Quantum
gravity.}

\pacs{04.60.-m, 04.80.Cc}

\maketitle

\section{Introduction}

The Galileo Galilei's Equivalence Principle between gravitational
and inertial masses in a combination with the local Lorentz
invariance of spacetime amount of the so-called Einstein's
Equivalence Principle. It is known to be a keystone of the
classical General Relativity [1,2]. Validity of this principle for
ordinary matter has been established so far with great accuracy,
$\frac{|m_i-m_g|}{m_i} \leq 10^{-17}-10^{-16}$, in the recent
space mission "MICROSCOPE" (see Refs. [3,4]), where $m_g$ and
$m_i$ are gravitational and inertial masses, respectively. In
literature, there are widely discussed possible new space
missions, "Galileo Galilei" [5] and "STEP" [6], which may increase
the above mentioned accuracy up to $\frac{|m_i-m_g|}{m_i} \leq
10^{-19}$.

The quantum theory of gravity has not been developed yet, but the
numerous speculation on this topic predict that the Einstein's
Equivalence Principle may be broken at extremely high energies, $E
\sim 10^{28} eV$, which will never be accessible for our
experimental studies. Nevertheless, recently we have shown (see
Refs.[7-9]) that even semiclassical variant of General Relativity,
where the field is not quantized but the matter is quantized,
predicts breakdown of the Einstein's Equivalence Principle at low
enough experimentally accessible energies. In particular, we have
shown [7,9] that the Einstein's Equivalence principle is broken
for passive gravitational mass at a microscopic level for a
composite quantum body. Indeed, electron in the hydrogen atom with
constant stationary energy, $E_n$, is not characterized by
constant passive gravitational mass [7-9]. According to the above
cited works, there exists a small (but non-zero) probability that
the quantum measurement of the electron mass gives the value $m_g
\neq m_e + \frac{E_n}{c^2}$. The situation with active
gravitational mass is , as shown [8,9],is even more interesting
since it breaks the Einstein's Equivalence Principle even at a
macroscopic level. Indeed, in the above mentioned papers, the
Equivalence Principle is considered for different macroscopic
ensembles of the hydrogen atoms. It is shown to survive for the
majority of the quantum ensembles, with the important exceptions,
which are macroscopic ensembles of the coherent quantum
superpositions of the stationary electron states, which we call
Gravitational demons. For such states, as demonstrated [8,9], the
expectation values of the mass can oscillate with time even in the
case where the expectation values of energy are constant.

\section{Goal}

In Sec. 4, we discuss in detail the breakdown of the Einstein's
Equivalence Principle between passive gravitational mass and
energy (i.e., inertial mass) at a macroscopic level, which was
first suggested by us in Ref. [9]. We show that the accepted by
majority of physicists accuracy of the validity of the Equivalence
Principle, $\frac{|m_i-m_g|}{m_i} \leq 10^{-17}-10^{-16}$, is
overestimated since they experimentally studied only usual
condensed matter samples. Below, we discuss behavior of several
different macroscopic ensembles of the simplest quantum composite
bodies - hydrogen atoms. In agreement with the above mentioned
experiments, we show that the Einstein's Equivalence Principle is
valid for almost all of them (see Sec. 3). We demonstrate that
this equivalence survives for macroscopic ensembles of the
stationary quantum electron states in a hydrogen atom due to the
so-called quantum virial theorem [10]. On the other hand, we
construct such quantum ensemble, which we call "Gravitational
demon" (in analogy with "Maxwell demon"), which breaks the
Einstein's Equivalence Principle at a macroscopic level [9]. The
Gravitational demon, by definition, is a a macroscopic ensemble of
the coherent superpositions of the stationary quantum states,
where the above mentioned equivalence is not survived due to the
the so-called quantum virial term.

Note that the virial term was first suggested in Ref.[11] (see
also Refs. [12] and [13]) for the classical model of a hydrogen
atom. In particular, it was shown that an external gravitational
field is coupled not with the total energy, $E = K+P$, but with
the following combination: $E +V$, where the virial term,
$V=2K+P$, with $K$ and $P$ being kinetic and potential energies,
respectively. Nevertheless, it was claimed [12,13] that the virial
term disappears, if we choose the local proper coordinates in the
gravitational field. Therefore, we suggest in the review two
methods to calculate passive gravitational mass: one, using
gravitational field as a perturbation in the Minkowski's metric
[9] (see Sub-sec. 4.1), and another one - using the local proper
coordinates (see Sub-sec. 4.2). We show that both methods for
macroscopic ensembles of the coherent superpositions of quantum
states in hydrogen atoms (i.e., for Gravitational demons) give the
same result - the breakdown of the Einstein's Equivalence
Principle [9]. In Sec. 5, some experimental aspects of the above
mentioned breakdown of the Equivalence Principle are discussed. In
particular, we pay attention that it is not necessary to conduct
very expensive experiments in space. It is possible to create the
Gravitational demons in the Earth's laboratories and, thus, to
discover the breakdown of the Einstein's Equivalence Principle in
the laboratories. Although the experiments are expected to be
rather difficult, the effect of the breakdown of the Equivalence
Principle may be very large and, in principle, even may be of the
order of unity.

\section{Einstein's Equivalence Principle for the stationary quantum states}

For further calculations, we use the textbook weak field
approximation [1,2] to describe spacetime outside the isotropic
gravitating body (e.g., the Earth),
\begin{equation}
ds^2 = -\biggl(1 + \frac{2\phi}{c^2} \biggl)(cdt)^2 + \biggl(1 -
\frac{2 \phi}{c^2} \biggl) (dx^2 +dy^2+dz^2 ), \ \ \phi = -
\frac{GM}{R},
\end{equation}
where $c$ is the velocity of light, $G$ is the gravitational
constant, $M$ is the Earth's mass, and $R$ is a distance from
center of the Earth. Then, in accordance with the Einstein's
Equivalence principle (which includes the local Lorentz
invariance), we can introduce the local proper spacetime
coordinates,
\begin{equation}
x'=\biggl(1-\frac{\phi}{c^2} \biggl) x, \ y'=
\biggl(1-\frac{\phi}{c^2} \biggl) y, \
z'=\biggl(1-\frac{\phi}{c^2} \biggl) z , \ t'=
\biggl(1+\frac{\phi}{c^2} \biggl) t,
\end{equation}
where spacetime is the Minkowski's one:
\begin{equation}
(ds')^2 = -(cdt')^2 + [(dx')^2 + (dy')^2 + (dz')^2].
\end{equation}
In the local proper spacetime coordinates (2), the Schr\"{o}dinger
equation for the electron wave functions in a hydrogen atom can be
approximately written in the following standard form:
\begin{equation}
i \hbar \frac{\partial \Psi({\bf r'},t')}{\partial t'} = \hat H
(\hat {\bf p'},{\bf r'}) \Psi({\bf r'},t')  ,
\end{equation}
with $\hat H (\hat {\bf p'},{\bf r'})$ being the standard
Hamiltonian for a hydrogen atom. It is important that, in Eq.(4)
and below, we disregard all the so-called tidal effects. In other
words, we consider the atom as a point-like body and do not
differentiate the gravitational potential with respect to the
relative electron coordinates, ${\bf r}$ and ${\bf r'}$. It is
easy to demonstrate that the disregarded tidal terms in the
electron Hamiltonian (4) are very small and are of the relative
order of $(r_B/R_0)|\phi/c^2| \sim 10^{-17}|\phi/c^2| \sim
10^{-26}$ in the Earth's gravitational field. [In this estimation,
$r_B$ is the typical "size" of the hydrogen atom (i.e., the Bohr's
radius) and $R_0$ is the Earth's radius.]

\subsection{Non-relativistic case}

Let us first consider the most important and principle case, where
we take account only the kinetic and Coulomb potential energies in
the the non-relativistic Schr\"{o}dinger equation for electron
wave functions in a hydrogen atom:
\begin{equation}
i \hbar \frac{\partial \Psi({\bf r'},t')}{\partial t'} = \hat H_0
(\hat {\bf p'},{\bf r'}) \Psi({\bf r'},t')  , \ \
 \ \hat H_0 (\hat {\bf p'},{\bf r'}) = m_e c^2 + \frac{\hat {\bf p'}^2}{2m_e}
 -\frac{e^2}{r'} .
\end{equation}
[In Eq.(5), as usual, $e$ is the electron charge, $r'$ is a
distance between electron and proton, and $\hat {\bf p'} = - i
\hbar
\partial /\partial {\bf r'}$ is electron momentum operator in the
local proper spacetime coordinates.] Below, we consider inertial
coordinate system, associated with the spacetime coordinates $(t,
x,y,z)$ in Eq.(2) and treat the weak gravitational field (1) as a
perturbation. As a result, we obtain the following Hamiltonian
[7,9]:
\begin{equation}
\hat H_0(\hat {\bf p},{\bf r}) = m_e c^2 + \frac{\hat {\bf
p}^2}{2m_e}-\frac{e^2}{r} + m_e  \phi + \biggl( 3 \frac{\hat {\bf
p}^2}{2 m_e} -2\frac{e^2}{r} \biggl) \frac{\phi}{c^2}.
\end{equation}
From Eq.(6), we rewrite the Hamiltonian in the following more
convenient form:
\begin{equation}
\hat H_0(\hat {\bf p},{\bf r}) = m_e c^2 + \frac{\hat {\bf
p}^2}{2m_e} -\frac{e^2}{r} + \hat m^g_e \phi \ ,
\end{equation}
where the passive gravitational mass operator of electron, $\hat
m^g_e$, is introduced by the equation:
\begin{equation}
\hat m^g_e  = m_e  + \biggl(\frac{\hat {\bf p}^2}{2m_e}
-\frac{e^2}{r}\biggl)/ c^2 + \biggl(2 \frac{\hat {\bf
p}^2}{2m_e}-\frac{e^2}{r} \biggl)/ c^2 \ ,
\end{equation}
which is equal to electron's weight operator in the weak field
(1). It is important that, in Eq.(8), only the first term
corresponds to the bare electron mass, $m_e$. To the bare electron
mass, there exist two corrections: the expected second term, which
corresponds to the electron energy contribution, and the
non-trivial third term, which is the so-called virial contribution
to the passive gravitational mass operator. As shown in Ref.[9],
Eqs.(7) and (8) can be directly obtained from the Dirac equation
in a weekly curved spacetime (1) [see, for example, Eq.(3.24) in
Ref. [14], if we disregard all tidal terms.

Let us discuss one important consequence of Eqs.(7) and (8). It is
easy to prove that the operator (8) does not commute with the
electron energy operator, taken in the absence of the field (1).
Therefore, from the beginning, it seems that the equivalence
between electron passive gravitational mass and its energy is
broken even for macroscopic ensemble of stationary quantum states.
But we show here that it is not so. In particular, to demonstrate
this equivalence at a macroscopic level, we consider a macroscopic
ensemble of hydrogen atoms being in a stationary quantum state
with a definite energy $E_n$. In this case, we calculate the
expectation value of the electron passive gravitational mass
operator (per unit atom) from Eq.(8) in the following way:
\begin{equation}
<\hat m^g_e > = m_e + \frac{ E_n}{c^2}  + \biggl< 2 \frac{\hat
{\bf p}^2}{2m_e}-\frac{e^2}{r} \biggl> /c^2 = m_e +
\frac{E_n}{c^2} .
\end{equation}
[Note that the third (virial) term in Eq.(9) is zero, according to
the so-called quantum virial theorem [10].] As a result of the
caculations, using the quantum virial theorem, we can conclude
that the equivalence between passive gravitational mass and energy
survives at a macroscopic level for stationary quantum states in
the non-relativistic approximation. We stress the important
difference between our quantum result [7,9] of Eq.(9) and the
corresponding result in classical case [12] is that the
expectation value of the passive gravitational mass corresponds to
averaging procedure over a macroscopic ensemble of the atoms,
whereas, in classical case, one averages over time.

\subsection{Relativistic corrections}

In this Subsection, we introduce a more general Hamiltonian, which
takes into account the so-called relativistic corrections to
electron wave functions in a hydrogen atom. It is well known [15],
that there are three correction terms, which can be derived from
relativistic Dirac equation and which have different physical
meanings. As a result, the total relativistic Hamiltonian in the
absence of gravitational field can be represented as:
\begin{equation}
\hat H (\hat {\bf p},{\bf r}) = \hat H_0 ( \hat {\bf p},{\bf r}) +
\hat H_1 (\hat {\bf p},{\bf r}) ,
\end{equation}
with the following corrections,
\begin{equation}
\hat H_1 (\hat {\bf p},{\bf r}) = \alpha \hat {\bf p}^4 + \beta
\delta^3 ({\bf r}) + \gamma \frac{\hat {\bf S} \cdot \hat {\bf L}
}{r^3} ,
\end{equation}
where the parameters $\alpha$, $\beta$, and $\gamma$ are:
\begin{equation}
\alpha = - \frac{1}{8 m_e^3 c^2}, \ \beta=\frac{\pi e^2
\hbar^2}{2m_e^2c^2}, \ \gamma = \frac{e^2}{2 m_e^2 c^2}.
\end{equation}
Let us discuss here the physical meaning of the relativistic
corrections. Note that the first contribution in Eq.(11) is called
the kinetic term, which follows from the relativistic relation
between energy and momentum. In the second correction, which has a
complicated physical meaning and is called Darwin's term,
$\delta^3 ({\bf r}) = \delta (x) \delta (y) \delta (z)$ is a three
dimensional Dirac's delta-function. And finally, the third
relativistic correction is the spin-orbital interaction, where
$\hat {\bf L} = - i \hbar [{\bf r} \times
\partial / \partial {\bf r}]$ is electron angular momentum
operator. Now, in the weak gravitational field (1), the
Schr\"{o}dinger equation for electron wave functions in the local
proper spacetime coordinates (2), in the absence of all tidal
effects, can be approximately written as:
\begin{equation}
i \hbar \frac{\partial \Psi({\bf r'},t')}{\partial t'} = [\hat
H_0(\hat {\bf p'},{\bf r'})+ \hat H_1(\hat {\bf p'},{\bf r'})]
\Psi ({\bf r',t'}).
\end{equation}

Using the coordinates transformation (2), the corresponding
relativistic Hamiltonian in the inertial coordinate system
$(t,x,y,z)$ can be expressed as:
\begin{eqnarray}
&\hat H(\hat {\bf p},{\bf r})= [\hat H_0(\hat {\bf p}, {\bf r}) +
\hat H_1 (\hat {\bf p},{\bf r})] \biggl(1 + \frac{\phi}{c^2}
\biggl)
\nonumber\\
&+\biggl(2 \frac{\hat {\bf p}^2}{2 m_e}-\frac{e^2}{r} + 4 \alpha
\hat {\bf p}^4 + 3 \beta \delta^3({\bf r}) + 3 \gamma \frac{\hat
{\bf S} \cdot \hat {\bf L} }{r^3} \biggl) \frac{\phi}{c^2} .
\end{eqnarray}
The operator of passive gravitational mass of electron, for the
relativistic Hamiltonian (14), can be written in more complicated
form than that in Eq.(8):
\begin{eqnarray}
&\hat m^g_e = m_e + \biggl( \frac{\hat {\bf p}^2}{2m_e} -
\frac{e^2}{r} + \alpha \hat {\bf p}^4 + \beta \delta^3 ({\bf r}) +
\gamma \frac{\hat {\bf S} \cdot \hat {\bf L} }{r^3} \biggl)/c^2
\nonumber\\
&+ \biggl(2 \frac{\hat {\bf p}^2}{2 m_e}-\frac{e^2}{r} + 4 \alpha
\hat {\bf p}^4 + 3 \beta \delta^3 ({\bf r}) + 3 \gamma \frac{\hat
{\bf S} \cdot \hat {\bf L} }{r^3} \biggl)/c^2 .
\end{eqnarray}
Here, we consider one more time a macroscopic ensemble of the
hydrogen atoms, with each of them being in a stationary quantum
state with a definite energy $E'_n$. Note that $E'_n$ takes into
account the relativistic corrections (11) to electron energy. In
this case, the expectation value of the relativistic electron mass
operator (15) per atom is:
\begin{equation}
<\hat m^g_e > = m_e + \frac{ E'_n}{c^2} + \biggl<2 \frac{\hat {\bf
p}^2}{2 m_e}-\frac{e^2}{r} + 4 \alpha \hat {\bf p}^4 + 3 \beta
\delta^3 ({\bf r}) + 3 \gamma \frac{\hat {\bf S} \cdot \hat {\bf
L} }{r^3} \biggl>/c^2 .
\end{equation}

We stress that the Einstein's Equivalence Principle will survive
at a macroscopic level if the expectation value of the third
(virial) term in Eq.(16) is zero. Below, we demonstrate,
therefore, that the Einstein's equation, related the expectation
value of passive gravitational mass and energy, can be applied to
the stationary quantum states. To this end, we define the
so-called virial operator [10],
\begin{equation}
\hat G = \frac{1}{2} (\hat {\bf p}{\bf r} +{\bf r} \hat {\bf p}) ,
\end{equation}
and make use of the standard equation of motion for its
expectation value:
\begin{equation}
\frac{d}{dt} \biggl< \hat G \biggl> = \frac{i}{\hbar} \biggl<
[\hat H_0(\hat {\bf p},{\bf r}) + H_1(\hat {\bf p},{\bf r}), \hat
G] \biggl> ,
\end{equation}
where $[\hat A, \hat B]$ denotes a commutator of two operators,
$\hat A$ and $\hat B$. Note that, in Eq.(18), the derivative
$d<\hat G>/dt$ has to be zero, since we consider the stationary
quantum state with a definite energy, $E'_n$. Therefore,
\begin{equation}
\biggl< [\hat H_0(\hat {\bf p},{\bf r})+ H_1(\hat {\bf p},{\bf
r}), \hat G] \biggl> = 0 ,
\end{equation}
where the Hamiltonian $\hat H_0(\hat {\bf p},{\bf r}) + \hat
H_1(\hat {\bf p},{\bf r})$ is defined by Eq.(13). Now, using
rather lengthy but straightforward calculations, we show that
\begin{eqnarray}
&&\frac{[\hat H_0(\hat {\bf p},{\bf r}), \hat G]}{-i \hbar}= 2
\frac{\hat {\bf p}^2}{2m_e}-\frac{e^2}{r} , \ \ \frac{[\alpha \hat
{\bf p}^4,\hat G]}{-i \hbar} = 4 \alpha \hat {\bf p}^4 ,
\nonumber\\
&&\frac{[\beta \delta^3 ({\bf r}), \hat G]}{-i \hbar} = 3 \beta
\delta^3 ({\bf r}), \ \ \frac{1}{-i \hbar} \biggl[ \gamma
\frac{\hat {\bf S} \cdot \hat {\bf L}}{r^3} , \hat G \biggl] = 3
\gamma \frac{\hat {\bf S} \cdot \hat {\bf L}}{r^3} ,
\end{eqnarray}
where we take into account the following equality:
\begin{equation}
x_i \frac{d [ \delta(x_i)]}{d x_i} = - \delta (x_i).
\end{equation}
From Eqs.(19) and (20), it is directly follows that
\begin{equation}
\biggl<2 \frac{\hat {\bf p}^2}{2 m_e}-\frac{e^2}{r} + 4 \alpha
\hat {\bf p}^4 + 3 \beta \delta^3 ({\bf r}) + 3 \gamma \frac{\hat
{\bf S} \cdot \hat {\bf L} }{r^3} \biggl> =0,
\end{equation}
and, thus, Eq.(16) can be represented in the Einstein's form:
\begin{equation}
<\hat m^g_e > = m_e + \frac{E'_n}{c^2} .
\end{equation}
Let us discuss the status of the Einstein's Equivalence Principle
for the stationary quantum states, considered in this Section.
Note that Eq.(23) directly establishes the equivalence between the
expectation value of electron passive gravitational mass and its
energy in a hydrogen atom, including the relativistic corrections.
Therefore, we can say that the Equivalence Principle survives for
quantum macroscopic bodies, which contains quantum composite
bodies in the stationary quantum states, which first shown in
Refs.[7] and [9]. On the other hand, for stationary quantum states
the Einstein's Equivalence Principle is broken at microscopic
level. Indeed, as first shown in Refs. [7] and [9], the quantum
measurement of passive gravitational mass in state with a definite
energy, $E_n$, can give with small probability the value $m^g_e
\neq m_e+\frac{E_n}{c^2}$. Although, we consider above and below
the simplest quantum composite body - a hydrogen atom, we
speculate that our results survive also for more complicated
quantum systems, including many-body systems with arbitrary
interactions of particles. These and further results reveal and
establish the physical meaning of a coupling of a macroscopic
quantum test body with a weak gravitational field.

\section{Inequivalence between passive gravitational mass and energy
at a macroscopic level} In this section, we discuss the so-called
Gravitational demon - such a macroscopic ensemble of quantum
states, which breaks the Einstein's Equivalence Principle. We show
that the Gravitational demon can be created, for example, as a
macroscopic ensemble of the coherent superpositions of two quantum
states in a hydrogen atom. To demonstrate that the expectation
value of gravitational mass of the above mentioned ensemble is not
equivalent to the expectation value of its energy, we use below
two different methods. In Sub-sec. 4.1, we apply the traditional
time-dependent quantum mechanical perturbation method [10] to
calculate electron wave functions in suddenly switched on
gravitational field and to show that the calculated energy in
gravitational field contains the unexpected
contribution\cite{Lebed-3} from the virial term. In Sub-sec. 4.2,
we use consideration of the problem in the local proper
coordinates (2) in gravitational field (1) and derive the same
unexpected virial contribution to electron gravitational mass, in
contract to classical physics, where the virial term disappear in
the proper local coordinates [12,13]. In the same way, as above,
in this Section, we disregard small probabilities of the order of
$\frac{\phi^2}{c^2}$ [see Refs.[7] and [9]] and, thus, ignore mass
quantization phenomenon.

\subsection{Inequivalence between passive gravitational mass and
energy for a macroscopic ensemble of the coherent superpositions
of two stationary quantum states (the first method)}

Let us first discuss the accepted in this review procedure of the
quantum measurements of passive gravitational mass for a
macroscopic ensemble. It is obvious that the expectation values of
energy and gravitational mass have to be calculated at the same
moment of time, $t=t'=0$. We suggest that, in the beginning (i.e.,
at $t<0$), where the gravitational field is absent, we have a
macroscopic ensemble of coherent superpositions of two wave
functions, corresponding to the ground state ($1S$) wave function,
$\Psi_1(r)$, and the first excited energy level ($2S$) wave
function, $\Psi_2(r)$, in a hydrogen atom:
\begin{equation}
\Psi(r,t) = \frac{1}{\sqrt{2}}\exp \biggl( \frac{-i m_e c^2 t}{
\hbar} \biggl)\biggl[ \exp \biggl(\frac{-i E_1 t}{ \hbar} \biggl)
\Psi_1(r) + \exp \biggl(\frac{-i E_2 t}{ \hbar} \biggl)
\Psi_2(r)\biggl] ,
\end{equation}
where we omit the $2P$ wave function since the wave functions with
different parity do not mix in the gravitational field (1) [7,9].
[It is important that a macroscopic coherent ensemble of such wave
functions, where the difference between phases of functions
$\Psi_1(r)$ and $\Psi_2(r)$ is fixed, is difficult but possible to
create by means of some laser technique [16].]Note that the
expectation value of energy in a macroscopic ensemble (24) in the
absence of gravitational field is equal to
\begin{equation}
<E (t<0)> = \frac{(E_1+E_2)}{2}.
\end{equation}

We perform the  following Gedanken experiment: we suddenly switch
on the gravitational field (1) at $t \geq 0$ [see the
corresponding perturbation (7) and (8) to the free electron
Hamiltonian, $\hat H_0$]. In this case, we can write the following
time-dependent electron Hamiltonian in the field:
\begin{equation}
U_1({\bf r},t) = \frac{\phi}{c^2}[m_e c^2 + \hat H_0({\bf r})
+\hat V({\bf r})] \Theta(t),
\end{equation}
where $\Theta(t)$ is the step-function and, as shown in Eq.(8),
the virial term can be represent as
\begin{equation}
\hat V({\bf r}) = 2 \frac{\hat p^2}{2m} - \frac{e^2}{r}.
\end{equation}

 As we have already mentioned, we disregard all small probabilities
 of the order of $\frac{\phi^2}{c^4}$ for electron to be in the gravitational
 field (1) in energy level with $n>2$ (see also Ref. [7] and [9]).
 Therefore, in a hydrogen atom we can consider only two levels
 with $n=1$ and $n=2$ and apply to them the two-level variant of the
 time-dependent perturbation theory [10]. According to this variant,
 in the gravitational field (1), the wave function can be represented as:
\begin{eqnarray}
\Psi^1(r,t) = \exp \biggl( \frac{-i m_e c^2 t}{ \hbar}
\biggl)\biggl[ &&\exp \biggl(\frac{-i E_1 t}{ \hbar} \biggl)
a_1(t) \Psi_1(r)
\nonumber\\
&&+ \exp \biggl(\frac{-i E_2 t}{ \hbar} \biggl) a_2(t)
\Psi_2(r)\biggl] .
\end{eqnarray}
By means of the standard quantum time-dependent perturbation
theory, we can obtain the following equations to determine the
functions $a_1(t)$ and $a_2(t)$ in Eq.(28):
\begin{eqnarray}
&&\frac{da_1(t)}{dt} = - i \ U_{11}(t)\ a_1(t) -i \ U_{12}(t) \exp
\biggl[-i\frac{(E_2-E_1) t}{\hbar} \biggl] \ a_2(t),
\nonumber\\
&&\frac{da_2(t)}{dt} = - i \ U_{22}(t)\ a_2(t) -i \ U_{21}(t) \exp
\biggl[-i\frac{(E_1-E_2) t}{\hbar} \biggl] \ a_1(t),
\end{eqnarray}
with the matrix elements of the perturbation (26),(27) being:
\begin{eqnarray}
&&U_{11}(t)= \Theta(t)\frac{\phi}{c^2} \int \Psi^*_1(r)[m_ec^2 +
\hat H_0({\bf r}) + \hat V({\bf r})] \Psi_1(r) d^3{\bf
r}=\Theta(t)\frac{\phi}{c^2}(m_ec^2+E_1),
\nonumber\\
&&U_{12}(t)= \Theta(t)\frac{\phi}{c^2} \int \Psi^*_1(r)[m_ec^2 +
\hat H_0({\bf r}) + \hat V({\bf r})] \Psi_2(r) d^3{\bf
r}=\Theta(t)\frac{\phi}{c^2} V_{12},
\nonumber\\
&&U_{22}(t)= \Theta(t)\frac{\phi}{c^2} \int \Psi^*_2(r)[m_ec^2 +
\hat H_0({\bf r}) +\hat V({\bf r})] \Psi_2(r) d^3{\bf
r}=\Theta(t)\frac{\phi}{c^2}(m_ec^2+E_2),
\nonumber\\
&&U_{21}(t)= \Theta(t)\frac{\phi}{c^2} \int \Psi^*_2(r)[m_ec^2 +
\hat H_0({\bf r}) + \hat V({\bf r})] \Psi_1(r) d^3{\bf
r}=\Theta(t)\frac{\phi}{c^2} V_{21},
\end{eqnarray}
where $V_{ij}$ are the matrix elements of the virial operator
(27). After solving Eqs.(29) and (30), it possible to find that
the function (28) is
\begin{equation}
\Psi^1(r,t) = \exp \biggl( \frac{-i m_e c^2 t}{ \hbar} \biggl)
\biggl[ \Psi^1_{1}(r,t) + \Psi^1_{2}(r,t)\biggl],
\end{equation}
where
\begin{eqnarray}
\Psi^1_{1}(r,t) = \frac{1}{\sqrt{2}} \exp \biggl[ -i \frac{(m_e
c^2 + E_1) \phi \ t}{c^2\hbar}&&\biggl]\exp \biggl(-i \frac{E_1
t}{\hbar} \bigg) \biggl[1 -\frac{\phi V_{12}}{c^2(E_2-E_1)}
\biggl] \Psi_1(r)
\nonumber\\
&&+\frac{1}{\sqrt{2}} \exp \biggl(-i \frac{E_2 t}{\hbar}
\bigg)\frac{\phi V_{12}}{c^2(E_2-E_1)}  \Psi_1(r)
\end{eqnarray}
and
\begin{eqnarray}
\Psi^1_{2}(r,t) = \frac{1}{\sqrt{2}} \exp \biggl[ -i \frac{(m_e
c^2 + E_2) \phi \ t}{c^2\hbar}&&\biggl] \exp \biggl(-i \frac{E_2
t}{\hbar} \bigg) \biggl[1 -\frac{\phi V_{21}}{c^2(E_1-E_2)}
\biggl] \Psi_2(r)
\nonumber\\
&&+\frac{1}{\sqrt{2}} \exp \biggl(-i \frac{E_1 t}{\hbar}
\bigg)\frac{\phi V_{21}}{c^2(E_1-E_2)}  \Psi_2(r).
\end{eqnarray}

It is easy to show that with accuracy to the first order of the
small parameter, $\frac{|\phi|}{c^2} \ll 1$, the wave function
(31)-(33) can be written in the following more convenient way:
\begin{eqnarray}
\Psi^1(r,t) = \frac{1}{\sqrt{2}} \exp \biggl[ -i \frac{(m_e c^2 +
E_1) (1+\phi/c^2) t}{\hbar}\biggl] \biggl\{&&\biggl[1 -\frac{\phi
V_{12}}{c^2(E_2-E_1)} \biggl] \Psi_1(r)
\nonumber\\
&&+\frac{\phi V_{21}}{c^2(E_1-E_2)}  \Psi_2(r) \biggl\}
\nonumber\\
+\frac{1}{\sqrt{2}} \exp \biggl[ -i \frac{(m_e c^2 + E_2)
(1+\phi/c^2) t}{\hbar}\biggl] \biggl\{&&\biggl[1 -\frac{\phi
V_{21}}{c^2(E_1-E_2)} \biggl] \Psi_2(r)
\nonumber\\
&&+\frac{\phi V_{12}}{c^2(E_2-E_1)}  \Psi_1(r) \biggl\},
\end{eqnarray}
where the wave function (34), taken with the same accuracy, is
normalized:
\begin{equation}
\int [\Psi^1(r,t)]^*\Psi^1(r,t) d^3r = 1 + O
\biggl(\frac{\phi^2}{c^4} \biggl).
\end{equation}

As we wrote before, wave functions in a macroscopic ensemble of
the quantum coherent superpositions at $t=t'=0$ [see Eq.(24)] are
characterized by the constant phase difference and, therefore, can
be written as
\begin{eqnarray}
\Psi(r,t) = \frac{1}{\sqrt{2}}\exp \biggl( \frac{-i m_e c^2 t}{
\hbar} \biggl)&&\biggl[ \exp \biggl(\frac{-i E_1 t}{ \hbar}
\biggl) \Psi^0_1(r)
\nonumber\\
&&+ \exp(i \tilde \alpha) \exp \biggl(\frac{-i E_2 t}{ \hbar}
\biggl) \Psi^0_2(r)\biggl] ,
\end{eqnarray}
with $ \Psi^0_1(r)$ and  $\Psi^0_2(r)$ being the corresponding
real functions and relative phase $\tilde \alpha$ being constant.
In this case, after some simple calculations, it is possible to
show that energy of the state (24) in the weak gravitational field
(1) is
\begin{eqnarray}
&&<E(t \geq 0)>= \int [\Psi^1(r,t)]^* \biggl(i \hbar
\frac{\partial}{\partial t} \biggl) \Psi^1(r,t) d^3r
\nonumber\\
&&=m_ec^2 \biggl(1 +\frac{\phi}{c^2}\biggl) +
\frac{(E_1+E_2)}{2}\biggl(1 +\frac{\phi}{c^2}\biggl)+ \tilde
V_{12}\frac{\phi}{c^2} \cos (\tilde \alpha) ,
\end{eqnarray}
where
\begin{equation}
\tilde V_{12} = \int \Psi^0_1(r) \hat  V({\bf r}) \Psi^0_2(r) d^3r
= \int
 \Psi^0_1(r) \biggl(2 \frac{\hat p^2}{2m} - \frac{e^2}{r} \biggl)
\Psi^0_2(r) d^3r.
\end{equation}
From Eq.(37), it is clear that macroscopic ensemble of the
coherent superpositions of quantum states (24),(36) (i.e.,
Gravitational demon) is characterized by the following expectation
value of electron mass per one hydrogen atom:
\begin{equation}
<m^g_e>= m_e + \frac{(E_1+E_2)}{2 c^2} + \frac{\tilde V_{12}}{c^2}
\cos (\tilde \alpha) ,
\end{equation}
where $\tilde \alpha=const$. Note that here $m_e$ is the bare
electron mass, the second term is the expected kinetic and
potential energy contributions to gravitational mass, whereas the
third virial term is non-trivial virial contribution to electron
mass. Therefore, Eq.(39) directly demonstrates inequivalence
between the expectation value of energy (25) and gravitational
mass of a macroscopic ensemble of the coherent superpositions of
the stationary states. Note that, if we have the incoherent
ensemble (where the phase $\tilde \alpha$ is not fixed), then the
gravitational mass (39) quickly oscillates with oscillating phase
and the equivalence between the expectation values (25) and (39)
restores. For the coherent macroscopic ensemble, Eq.(39) crucially
depends on the ensemble preparation procedure (i.e., on the phase
difference $\tilde \alpha$). For instance, the expectation value
of the gravitational mass can be both larger and smaller than the
expected value from the Einstein's Equivalence Principle and for
the simple cases $\tilde \alpha_{1} = 0$, $\tilde \alpha_2 =
\pi/2$, and $\tilde \alpha_{3} = \pi$ is equal:
\begin{equation}
<m^g_e>= m_e + \frac{(E_1+E_2)}{2 c^2} + \frac{\tilde
V_{12}}{c^2}, \ \ \tilde \alpha_1 =0,
\end{equation}
\begin{equation}
<m^g_e>= m_e + \frac{(E_1+E_2)}{2 c^2}, \ \ \tilde \alpha_2
=\pi/2,
\end{equation}
and
\begin{equation}
<m^g_e>= m_e + \frac{(E_1+E_2)}{2 c^2} - \frac{\tilde
V_{12}}{c^2}, \ \ \tilde \alpha_3 = \pi.
\end{equation}
To make sure that the suggested effect is not zero, we have
calculated the virial matrix element for a hydrogen atom and found
that
\begin{equation}
 \tilde V_{12} = 0.56 \ (E_2-E_1).
\end{equation}

\subsection{Inequivalence between passive gravitational mass and
energy for macroscopic ensemble of the coherent superpositions of
two stationary quantum states (the second method)}

Below, we make use of different method to obtain Eq.(39). As shown
in Refs. [12] and [13], in classical case the virial term
disappear if we introduce the proper local coordinates
$(x',y',z')$ [see Eq.(2)]. Therefore, it is important to show that
Eq.(39) survives in quantum case if we consider the problem in the
proper local coordinates (2). Here, to measure gravitational mass,
we perform the same Gedanken experiment as in Sub-sec. 4.1. In
particular, at $t<0$, there is no gravitational field and we have
a macroscopic ensemble of the coherent superposition of two
stationary wave functions [see Eq.(24)]. At $t=0$, we switch on
gravitational field (1), which is equivalent to a change of
geometry of the space. According, to the local Lorentz invariance,
which is a part of the Einstein's Equivalence Principle, the
solution of the Schr\"{o}dinger equation at $t \geq 0$ can be
written in the local proper coordinates (2) as
\begin{eqnarray}
\Psi_2(r',t') = \exp \biggl( \frac{-i m_e c^2 t'}{ \hbar}
\biggl)&\biggl[ A \exp \biggl(\frac{-i E_1 t'}{ \hbar} \biggl)
\Psi^0_1(r')
\nonumber\\
&+ B \exp \biggl(\frac{-i E_2 t'}{ \hbar} \biggl)
\Psi^0_2(r')\biggl] ,
\end{eqnarray}
where complex coefficients $A$ and $B$ are not necessarily equal
to $\frac{1}{\sqrt{2}}$ and take into account the gravitational
field (1). On the other hand, in the proper local coordinates, we
can represent the wave function (24) at $t=0$ as
\begin{eqnarray}
&&\Psi_3(r',t') = \frac{1}{\sqrt{2}}\exp \biggl[ \frac{-i m_e c^2
t'(1-\phi/c^2)}{ \hbar} \biggl] \biggl( 1+\frac{\phi}{c^2}
\biggl)^{3/2}
\nonumber\\
&&\times\biggl\{ \exp \biggl[\frac{-i E_1 t'(1-\phi/c^2)}{ \hbar}
\biggl] \Psi^0_1[r'(1+\phi/c^2)] \
\nonumber\\
&&+ \exp(i \tilde \alpha) \exp \biggl[\frac{-i E_2
t'(1-\phi/c^2)}{ \hbar} \biggl] \Psi^0_2[r'(1+\phi/c^2)]\biggl\} .
\end{eqnarray}
[Note that the wave function (44) is normalized in the proper
local coordinates (2).] It is important that, at $t=t'=0$, the
above discussed wave functions (44) and (45) have to be equal to
each other:
\begin{equation}
\Psi_2(r',t'=0) =\Psi_3(r',t'=0).
\end{equation}
Using Eq.(46) and the following orthogonality condition for real
functions $\Phi^0_1(r')$ and $\Phi^0_2(r')$,
\begin{equation}
\int^{\infty}_0 [\Psi^0_1(r')]^2 d^3r' = \int^{\infty}_0
[\Psi^0_2(r')]^2 d^3r' =1, \ \ \int^{\infty}_0 \Psi^0_1(r')
\Psi^0_2(r') d^3r'=0,
\end{equation}
it is possible to define the coefficients $A$ and $B$ in Eq.(44):
\begin{equation}
A= \frac{1}{\sqrt{2}}[\Delta_{11}+\exp(i\tilde
\alpha)\Delta_{12}], \ \ B= \frac{1}{\sqrt{2}}[\Delta_{21}+
\exp(i\tilde \alpha)\Delta_{22}],
\end{equation}
where
\begin{equation}
\Delta_{ij}= \biggl(1+ \frac{\phi}{c^2} \biggl)^{3/2}
\int^{\infty}_0 \Psi^0_i(r')
\Psi^0_j\biggl[r'\biggl(1+\frac{\phi}{c^2} \biggl)\biggl] d^3r'.
\end{equation}

Let us calculate the matrix elements of the matrix $\hat
\Delta_{ij}$ with the accepted in this review accuracy - to the
first order of the small parameter $|\frac{\phi}{c^2}| \ll 1$. By
the definition (49),
\begin{eqnarray}
\Delta_{11}= \biggl(1+ \frac{\phi}{c^2} \biggl)^{3/2}
\int^{\infty}_0 \Psi^0_1(r')
\Psi^0_1\biggl[r'\biggl(1+\frac{\phi}{c^2} \biggl)\biggl] d^3r'
\nonumber\\
\approx \biggl(1- \frac{\phi}{c^2} \biggl)^{3/2} \int^{\infty}_0
\Psi^0_1\biggl[r\biggl(1-\frac{\phi}{c^2} \biggl)\biggl]
\Psi^0_1(r) d^3r.
\end{eqnarray}
From Eq.(50), it follows that $\Delta_{11}$ is an even function of
the variable $\frac{\phi}{c^2}$, therefore, in our approximation
\begin{equation}
\Delta_{11} = \int_0^{\infty} [\Psi^0_1(r')]^2 d^3r' = 1.
\end{equation}
Using the same method, it is easy to show that
\begin{equation}
\Delta_{2,2} = \int_0^{\infty} [\Psi^0_2(r')]^2 d^3r' = 1.
\end{equation}
Calculations of non-diagonal matrix elements of the matrix $\hat
\Delta_{ij}$ is more complicated procedure. Let us start from
calculation of the matrix element $\Delta_{12}$. From definition
of the matrix (49), we have
\begin{eqnarray}
&&\Delta_{12}= \biggl(1+ \frac{\phi}{c^2} \biggl)^{3/2}
\int^{\infty}_0 \Psi^0_1(r')
\Psi^0_2\biggl[r'\biggl(1+\frac{\phi}{c^2} \biggl)\biggl] d^3r'
\nonumber\\
&&\approx \biggl( 1-\frac{\phi}{c^2} \biggl)^{3/2} \int^{\infty}_0
\Psi^0_1\biggl[r \biggl(1-\frac{\phi}{c^2} \biggl)\biggl]
\Psi^0_2(r) d^3r
\nonumber\\
&&\approx \biggl( 1-\frac{\phi}{c^2} \biggl)^{3/2} \int^{\infty}_0
\Psi^0_1(r) \Psi^0_2(r) d^3r -\frac{\phi}{c^2} \int_0^{\infty}
\Psi'_1(r) \ r \ \Psi^0_2(r) d^3r = -\frac{\phi}{c^2} \ \Delta,
\nonumber\\
\end{eqnarray}
where
\begin{equation}
\Psi'_1(r) = \frac{d\Psi^0_1}{dr}, \ \ \ \ \Delta =
\int_0^{\infty} \Psi'_1(r) \ r \ \Psi^0_2(r) d^3r.
\end{equation}
Using the same way, we can find that
\begin{equation}
\Delta_{21} = + \frac{\phi}{c^2} \ \Delta .
\end{equation}

For further development, it is necessary to calculate quantity
$\Delta$ in Eqs.(53)-(55) in terms of the virial term. In
particular, let us show that
\begin{equation}
\Delta= \int^{\infty}_0 \Psi'_1(r) r \Psi^0_2(r) d^3 r =
\frac{\tilde V_{12}}{E_2-E_1},
\end{equation}
where the matrix elements of the virial operator (27), $V_{1n}$,
are define by the standard equation:
\begin{equation}
\tilde V_{12}=\int^{\infty}_0 \Psi^0_1(r) \hat V(r) \Psi^0_2(r)
d^3r.
\end{equation}
To this end, we rewrite the Schr\"{o}dinger equation in
gravitational field (5) in terms of the unperturbed spacetime
coordinates $(t,x,y,z)$ (2):
\begin{eqnarray}
(m_e c^2 + E_1)
\Psi^0_1\biggl[\biggl(1-\frac{\phi}{c^2}\biggl)r\biggl] = \biggl[
m_e c^2&&- \frac{1}{(1-\phi/c^2)^2} \frac{\hbar^2}{2m} \biggl(
\frac{\partial^2}{
\partial x^2} + \frac{\partial^2}{ \partial y^2} +
\frac{\partial^2}{\partial z^2} \biggl)
\nonumber\\
&&- \frac{1}{(1-\phi/c^2)} \frac{e^2}{r} \biggl]
\Psi^0_1\biggl[\biggl(1-\frac{\phi}{c^2}\biggl)r\biggl] .
\end{eqnarray}
Then, using the accepted week field approximation and, thus,
keeping only terms of the first order with respect to the small
parameter $|\frac{\phi}{c^2}| \ll 1$, we obtain:
\begin{eqnarray}
E_1 \Psi^0_1(r) - \frac{\phi}{c^2} \ E_1 \ r \ \Psi'_1(r)=
&&\biggl[ - \frac{\hbar^2}{2m_e} \biggl( \frac{\partial^2}{
\partial x^2} + \frac{\partial^2}{ \partial y^2} +
\frac{\partial^2}{\partial z^2} \biggl) - \frac{e^2}{r} +
\frac{\phi}{c^2} \ \hat V(r) \biggl]
\nonumber\\
&&\times\biggl[ \Psi^0_1(r) - \frac{\phi}{c^2} \ r \ \Psi'_1(r)
\biggl].
\end{eqnarray}
As follows from Eq.(59),
\begin{equation}
- E_1 r  \Psi'_1(r) = \biggl[ \frac{\hbar^2}{2m_e} \biggl(
\frac{\partial^2}{
\partial x^2} + \frac{\partial^2}{ \partial y^2} +
\frac{\partial^2}{\partial z^2} \biggl) + \frac{e^2}{r} \biggl] r
\Psi'_1(r) +\hat V(r) \Psi^0_1 (r).
\end{equation}
Let us multiply Eq.(60) on $\Psi^0_2(r)$ and integrate,
\begin{eqnarray}
- E_1 \int^{\infty}_{0} \Psi^0_2(r) r \Psi'_1(r) d^3r&&=
\int^{\infty}_{0} \Psi^0_2(r) \biggl[ \frac{\hbar^2}{2m_e} \biggl(
\frac{\partial^2}{
\partial x^2} + \frac{\partial^2}{ \partial y^2} +
\frac{\partial^2}{\partial z^2} \biggl) + \frac{e^2}{r} \biggl] r
\Psi'_1(r)d^3r
\nonumber\\
&&+\int^{\infty}_{0} \Psi^0_2(r)\hat V(r) \Psi^0_1 (r)d^3r.
\end{eqnarray}
If we make use of the fact that the Hamiltonian and virial term
are the Hermitian operators, we can rewrite Eq.(61) as
\begin{eqnarray}
E_1 \int^{\infty}_{0} \Psi^0_2(r) r \Psi'_1(r) d^3r &&= E_2
\int^{\infty}_{0} \Psi^0_2(r) r \Psi'_1(r) d^3r
\nonumber\\
&&- \int^{\infty}_{0} \Psi^0_1(r)\hat V(r) \Psi^0_2(r)d^3r.
\end{eqnarray}
Then, Eq.(56) directly follows from Eq.(62):
\begin{equation}
\Delta = \int^{\infty}_{0}\Psi'_1(r)  r \Psi_2(r) d^3r =
\frac{\tilde V_{1,2}}{E_2-E_1}.
\end{equation}

Firstly, let us check that the wave function (44) is normalized in
the proper local coordinates (2) with the accepted accuracy of our
calculations. To this end, we calculate:
\begin{eqnarray}
&\int^{\infty}_0 |\Psi_2(0,r')|^2 d^3r' = \int^{\infty}_0 [A^*
\Psi^0_1(r') + B^* \Psi^0_2(r')] [A \Psi^0_1(r') + B \Psi^0_2(r')]
d^3r'
\nonumber\\
&=|A|^2 + |B|^2,
\end{eqnarray}
where
\begin{eqnarray}
|A|^2 + |B|^2 = \frac{1}{2} \biggl\{\biggl[1- \frac{\phi}{c^2}
\Delta \exp(-i \tilde \alpha)  \biggl] \biggl[1- \frac{\phi}{c^2}
\Delta \exp(+i \tilde \alpha) \biggl]
\nonumber\\
+\biggl[\exp(-i \tilde \alpha) + \frac{\phi}{c^2} \Delta \biggl]
\biggl[\exp(+i \tilde \alpha) + \frac{\phi}{c^2} \Delta
\biggl]\biggl\} \approx 1
\end{eqnarray}
[see Eqs.(48),(53), and (55)]. $\\$ Second, let us calculate the
expectation value of energy for wave function (44) in the
gravitational field (1):
\begin{eqnarray}
&&<E> = |A|^2 (E_1+m_ec^2) \biggl( 1+\frac{\phi}{c^2} \biggl) +
|B|^2 (E_2+m_ec^2) \biggl( 1+\frac{\phi}{c^2} \biggl)
\nonumber\\
&&=\frac{1}{2} \biggl\{(E_1+m_ec^2) \biggl(1 +\frac{\phi}{c^2}
\biggl) \biggl[1- \frac{\phi}{c^2} \Delta \exp(-i \tilde \alpha)
\biggl] \biggl[1- \frac{\phi}{c^2} \Delta \exp(+i \tilde \alpha)
\biggl]
\nonumber\\
&&+ (E_2+m_ec^2) \biggl( 1+\frac{\phi}{c^2} \biggl) \biggl[\exp(-i
\tilde \alpha) + \frac{\phi}{c^2} \Delta \biggl] \biggl[\exp(+i
\tilde \alpha) + \frac{\phi}{c^2} \Delta \biggl]\biggl\}
\nonumber\\
&&\approx m_ec^2 + \frac{1}{2}(E_1+E_2) + m_e \phi +
\frac{1}{2c^2}(E_1+E_2)\phi + \frac{\tilde V_{12}}{c^2}
\cos(\tilde \alpha) \phi .
\end{eqnarray}
The equation for the expectation value of gravitational electron
mass (39) directly follows from Eq.(66) and contains the discussed
above virial term. So in this Subsection, we have derived Eq.(39),
using the local proper coordinates (2), in contrast to the
classical case statements [12,13].

\section{Some experimental aspects.}

From our results, discussed in the review, it follows that in
order to discover violations of the Einstein's Equivalence
Principle it is not necessary to perform long and expensive
experiments in the space. What we actually need is to create
special macroscopic ensemble of the coherent superpositions of two
or several stationary quantum states (which we call Gravitational
demon) and measure its weight. Such macroscopic ensemble, it is
possible to create using laser technology (see Ref. [16]). Then,
it is necessary to measure the Gravitation demon's energy and
compare the above mentioned two quantities. To evaluate energy in
the considered case, it is a good idea just to count a number of
the emitted photons from the macroscopic ensemble of the atoms.
Let us discuss some obvious difficulties. It is evident that
weight of the created Gravitational demon has to be measured at
the moment of time which is very close to its creation. In the
review, we have used for introduction of the gravitational field
the so-called step-like function, $\Theta(t)$. Of course, this
does not mean a motion in the gravitational field (1) (or a motion
of the gravitational source) with speed higher than the speed of
light. We can use step-like function, if significant change of the
gravitational field happens quicker than the characteristic period
of quasiclassical rotation of electrons in a hydrogen atom. More
strictly speaking, in our case of superposition of two atomic
levels, we need the time about $\delta t \leq t_0 =\frac{2 \pi
\hbar}{E_2 -E_1} \sim 10^{-15}s$. To conclude, we have
demonstrated, using Gedanken experiments at the Earth's
laboratory, that breakdown of the Einstein's Equivalence Principle
is possible for some special macroscopic quantum states. Finding
the best concrete realizations for the corresponding real
experiments are the topic for the future investigations.

\section*{Acknowledgments}

We are thankful to Natalia N. Bagmet (Lebed), Steven Carlip,
Fulvio Melia, Pierre Meystre, Keneth Nordtvedt, Douglas Singleton,
Elias Vagenas, and Vladimir E. Zakharov for numerous useful
discussions.


\begin{thebibliography}{0}

\bibitem{Landau}
L.D. Landau and E.M. Lifshitz, {\it The Classical Theory of
Fields}, 4th edn. (Butterworth-Heinemann, 2003).

\bibitem{Misner}
C.W. Misner, K.S. Thorne, and J.A. Wheeler, {\it Gravitation}
(W.H. Freeman and Co, 1973).

\bibitem{Microscope-1}
Pirre Touboul, Gilles Metriss, Manuel Rodrigues et al., Phys. Rev.
Lett. {\bf 119}, 231101 (2017).


\bibitem{Microscope-2}
Pirre Touboul, Gilles Metriss, Manuel Rodrigues et al., Class.
Quantum Grav. {\bf 36}, 225006 (2019).

\bibitem{GG}
Anna M. Nobili and Alberto Anselmi, Phys. Rev. D {\bf 98}, 042002
(2018).

\bibitem{Step}
J.P. Pereira, J.M. Overduin, and A.J. Poyneer, Phys. Rev. Lett.
{\bf 117}, 071103 (2016).

\bibitem{Lebed-1} A.G. Lebed, {\it Adv. High Ener. Phys.} {\bf 2014},
678087 (2014).

\bibitem{Lebed-2} A.G. Lebed, {\it J. Phys.: Conf. Ser.} {\bf 738},
012036 (2016).

\bibitem{Lebed-3} A.G. Lebed, {\it Int. J. Mod. Phys. D} {\bf 28},
1930020 (2019).

\bibitem{Park}
D. Park, {\it Introduction to the Quantum Theory}, 3rd edn. (Dover
Publications, 2005).

\bibitem{Nord}
K. Nordtvedt, {\it Class. Quantum Grav.} \textbf{11}, A119 (1994).

\bibitem{Carlip}
S. Carlip, {\it Am. J. Phys.} \textbf{66}, 409 (1998).

\bibitem{Zych-1}
M. Zyck, L. Rudnicki, and I. Pikovski, arXiv:1808.05831v1 (2018).

\bibitem{Fisch}
E. Fischbach, B.S. Freeman and W.K. Cheng, {\it Phys. Rev. D}
\textbf{23}, (1981) 2157.

\bibitem{Schwabl}
See, for example, F. Schwabl, {\it Advanced Quantum Mechanics} 3rd
edn. (Springer, 2005).

\bibitem{Meystre}
Pierre Meystre, private communication (unpublished) (2020).



\end{thebibliography}
\end{document}